# AMBIGUITY INVOKES CREATIVITY: LOOKING THROUGH QUANTUM PHYSICS

Souparno Roy[*1,2], Archi Banerjee[1,2], Ranjan Sengupta[2], Dipak Ghosh[2]

[1]Department of Physics, Jadavpur University, Kolkata-700032

[2]Sir C.V. Raman Centre for Physics and Music, Jadavpur University, Kolkata- 700032

[*]thesouparnoroy@gmail.com

**ABSTRACT:**

*Creativity, defined as 'the tendency to generate or recognize new ideas or alternatives and to make connections between seemingly unrelated phenomena', is too vast a horizon to be summed up in such a simple sentence. The extreme abstractness of creativity makes it harder to quantify in its entirety. Yet, a lot of efforts have been made both by psychologists and neurobiologists to identify its signature. A general conformity is expressed in the 'Free association theory', i.e. the more freely a person's conceptual 'node's are connected, the more divergent thinker (also, creative) he or she is. Also, tolerance of ambiguity is found to be related to divergent thinking. In this study, we approach the problem of creativity from a theoretical physics standpoint. Theoretically, for the initial conceptual state, the next 'jump' to any other node is equally probable and non-deterministic. Repeated intervention of external stimulus (analogous to a 'measurement') is responsible for such 'jumps'. And to study such a non-deterministic system with continuous measurements, Quantum theory has been proven the most successful, time and again. We suggest that this collection of nodes form a system which is likely to be governed by quantum physics and specify the transformations which could help explain the conceptual jump between states. Our argument, from the point of view of physics is that the initial evolution of the 'creative process' is identical, person or field independent. To answer the next obvious question about individual creativity, we hypothesize that the quantum system, under continuous measurements (in the form of external stimuli) evolves with chaotic dynamics, hence separating a painter from a musician. Possible experimental methodology of these effects has also been suggested using ambiguous figures.*

**INTRODUCTION:**

"Others have seen what is and asked why. I have seen what could be and asked why not. " — Pablo Picasso.

From the very dawn of civilization, Creativity has been inspiring and reshaping human existence continuously. It was creativity that gave rise to the likes of Picasso, da Vinci, and Einstein- who, with their endless wonders on and off the paper, changed the course of human history and civilization time and again. We, in turn, have strived to understand the experiences of them and have questioned what, if anything, we ourselves have in common with these amazing individuals. Creativity is the development of new ideas and original products in a novel and appropriate way [1][2][3][4]. And theories and ideas about understanding the creative process stem from far back in history since it is a particularly human characteristic [5].

Though it started as far back as late 1800s, the systemic search of creativity blossomed in the twentieth century, where its roots have been searched in the lights of a plethora of diversified disciplines [6]:

- **Psychoanalytic approach**: Freud's discussion of creativity as the sublimation of drives [7], Winnicott's work on development which makes creativity central and intrinsic to human nature [8] [9] etc.

- **Cognitive approach:** Originated from Galton's work on hereditary genius [10]. Also includes Mednick's exploration of the associative process [11] and Guilford's exploration of divergent production of ideas and products [12][13].
- **Behaviourist approach:** B.F. Skinner's discussion of effect of chance mutation in behaviours [14][15][16].
- **Neurological/Biological approach:** using modern instruments like EEG or fMRI to pinpoint the brain areas activated during creativity (though not much conformity has been found) [17][18].

The latter half of the 20<sup>th</sup> century of creativity research is dominated by Psychometrics. Psychometric approaches to creativity were begun by psychologist J.P. Guilford, who developed a tool for measuring the extent of divergent thinking, which he later developed into the concept of 'divergent production'[13][19]. Divergent production (or thinking), also loosely called 'lateral thinking', is a method used to generate multiple related ideas for a given topic or a problem. Despite criticism, the idea of divergent thinking has become important in the scientific study of creativity because many widely used tests for creativity are measures of individual differences in divergent thinking ability like the Torrance tests of creative thinking [20][21][22].

**FREE ASSOCIATION THEORY OF CREATIVITY:**

Divergent thinking tasks have been widely used because traditionally creativity has been understood in terms of the accessibility of concepts in our long term memory systems. Concepts are connected in our brains in 'semantic networks'. Here is a schematic of a semantic network, with each concept 'node' of the network accessible from the concept 'street' via other nodes.

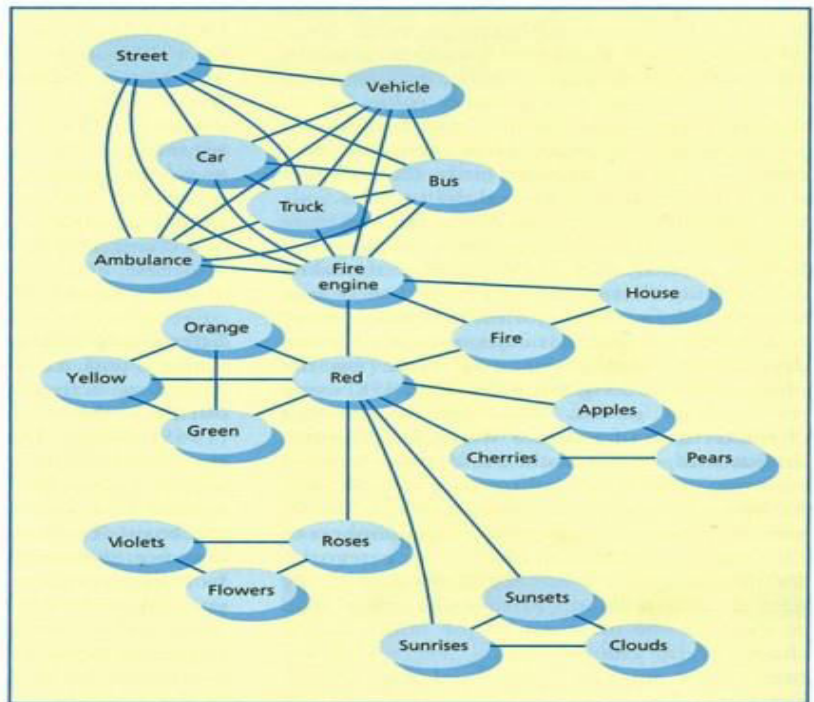

Figure 1: Schematic diagram of a semantic network [23]

Psychologists have proposed that individual differences in creativity are due to differences in whether these kinds of associative networks were 'steep' or 'flat' – those with 'flat' networks have numerous and loose conceptual connections, enabling them to be more creative. Those with 'steep' networks tend to have more logical, linear associations between nodes [11][24].

In this paper, we investigate this very idea of divergent thinking from a theoretical point of view of Physics, more specifically, Quantum physics.

- **Why use Physics? And why Quantum Physics?**:

Answer to the first question- why shouldn't we? Brains are nothing more, and nothing less, than atoms joined together to form molecules, bounded into specialized cells, i.e., neurons, which can communicate with each other. Like everything else in this world, neurons and brains must obey the laws of physics.

As for the second part of the question: at first sight it may seem bizarre, or even ridiculous, to draw a connection between creative process (or any such higher brain functions, for that matter), something lying within the realm of day-to-day human behavior – on the one hand and quantum mechanics – a highly successful theory devised mainly to explain microscopic subatomic phenomena on the other hand. Yet, there are good scientific reasons to do so [25]. Quantum theory, with astonishing counterintuitive ramifications, it is the best empirically confirmed scientific theory in human history. It is essential to every natural science and its practical applications, such as the laser and the transistor, have paved the way for new groundbreaking ideas. The application of Quantum theory to human cognition (and also, creativity) is driven not only by deep resonations between basic notions of quantum theory and psychological conceptions and intuitions, but also by the potential of the theory to provide coherent and mathematically principled explanations for the puzzles and challenges in human cognitive research. A very brief overview of the unusual nature of the theory is presented as a precursor to the idea we want to convey.

**Review of some prominent quantum phenomena and properties:**

- Discreteness of nature: Quantum physics emphasizes that our world is built on discrete particles that are bound in finite systems of discontinuous energies [26]. Unlike classical physics, where energies have continuous distributions.
  Also in the neurological framework, to describe the dynamics of neuron firings- evidently a discrete and discontinuous process- quantum theory can be used for more precision instead of existing classical ideas.

- Wave-particle dual nature: Particles, like electron or photon, can exhibit both particle and wave characteristics, an event that is entirely non-classical in nature [27].
  Proposal of the manifestation of this property in mind-matter relationship is not an alien notion [28].

- Quantum tunneling: Quantum wave effects allow tunneling through an energy barrier which would classically be insurmountable [29].
  It is suggested that macroscopic spread of quantum effects in human brain may involve the tunneling effect [30].

- Quantum superposition: Before a measurement, a particle can be in a state which is a superposition of all the possible energy configurations available for the particle [31].
  Dealing with a mental space which consists several possible states, the role of quantum superposition is surely undeniable.

- Indeterministic nature: Quantum process is indeterministic (i.e., the process of measurement introduces indeterminism). Unlike classical counterpart, the actual outcomes of an experiment are not uniquely determined by the theory [32]. Neurological investigation hasn't been able to determine the exact state of the psychological functions in any sort of experiments. Limitations of the deterministic ideas definitely point towards a quantum intervention.

- Quantum entanglement: Entanglement is the inseparable quantum correlation of two or more particles or degrees of freedom which determines the states of these two spatially separated systems simultaneously as soon as one of them is acted or measured upon [33]. Entanglement effects in human brain have been of a topic of several researches, including memory [34] or cognitive processes [35].

These are some of the unique phenomenon that marks the stark differences between quantum theory and classical physics, making the former more suitable to approach the hard problem of explaining the higher brain functions, including Creativity [36].

## QUANTUM LEAP INTERPRETATION:

Before we explain the hypothesis, let us steer the reader towards the path we hope to take. From the viewpoint of cognitive neurobiology, we now well understand the nature of nerve cell activity: the creation of action potentials, ion exchange, the use of energy, axonal transport, the vesicle cycle, and the production, cycle and breakdown of neurotransmitters. Yet, how these unconscious materials produce the stream of consciousness or perform the individual higher brain functions so smoothly, classical theories are still mum on that. Hence, we take help from the emerging field of Quantum neurobiology that explores these issues with an alternate approach. According to quantum neurobiology, Quantum physics is involved in biological processes, and consciousness, memory, internal experiences, and the processes of choice and decision making, which are the products of the warm-wet-noisy brain, may be the result of the operations of quantum physics [37]. To make our intention crystal clear, we would like to state that we never claim our interpretation to be the only valid one by which the creative process can be explained. Our argument is that, this point of view can help paving the way for a foundation to further understanding this complex brain process.

Keeping the above preface in mind, and also the Free association theory of creativity, we separate the whole 'creative process' in three distinct parts:

1. **Picking up stage**: Brain receives the external stimulus which constitutes the first 'node' in the creative process (from here the mental state, represented by a vector in a Hilbert space, will start its divergent evolution). This is equivalent to the situation where a creative person 'picks up' his inspiration.

2. **Leap stage**: The mental state, therefore, starts evolving in a 'steep' or 'loose' path to subsequent nodes by taking 'leap's between them. This is where the divergent thinking helps the 'node' to make distant nodal connections, expanding the creative process spatially and/or temporally.

3. **Chaotic Evolution stage**: With continuous measurement of the evolution function (measurement here signifies the continuous bombardment of contextual inputs from memory and/or the external stimulus), the evolution of the vector is encountered by a 'white noise', under whose influence, it exhibits nonlinearity and a chaotic nature. This is the phase, which separates two creative individuals (say a musician and a painter) depending on the nature of the memory inputs and stimulus received.

## ELABORATION OF THE STAGES:

## 1. **Picking up stage:**

We start by using the standard method to explain any dynamical system and its evolution. i.e., constructing a phase space. Phase space is a space which contains all possible states of a physical system. In this case, the physical system we are interested in can be represented by a 'Mental State function', a wave function on the 'Mental state space'. Using Dirac notation, we denote this state function as $|\Psi>$. Unlike the classical approach where state of a system is denoted by a specific point in phase space, in the quantum mechanical approach, this is a complex valued wave function whose position and momentum cannot be determined simultaneously (Heisenberg's uncertainty principle) [38]. The state function can evolve in time and can be changed by 'interactions' with external stimulus or memory (or experiences). Also, our constructed space is made up of all the probable states that this state function can achieve after these 'interactions'. So, to generalize, we can write |Ψ> as a linear combination of all these probable states, which serve as orthogonal basis of the Hilbert space of all mental states:

$$|\Psi> = \sum_{k=1}^{n} c_k |k>;$$ where $c_k$ 's are complex coefficients and $|k>$ are basis vectors which span the Hilbert space of all probable mental states.

Now, when an external stimulus interacts with the wave function, |Ψ> changes to a different state from its initial state. Manousakis [39] showed that this interaction happens via a concept known as operators. In quantum mechanical point of view, operators are the equivalent of observables in classical physics. So, when the state function interacts with environment, a particular operator, say $\hat{Q}$ (it has a matrix representation), operates on |Ψ> and changes it to a new state |Ψ'>. This is given as-

$$|\Psi'> = \hat{Q} \ |\Psi>$$

This state transition depends on the structure and properties of $\hat{Q}$. Manousakis [39] used the example of binocular rivalry and showed how the two possible state transitions are achieved. In our interpretation, the initial stimulus (which 'provokes' the creative process) operates in the same manner, culminating in the new state |Ψ'>. This |Ψ'> works as the first 'node' of the creative network. Hence, change in the Mental state function is what starts the creative process, i.e., the 'Picking up' stage.

2. **Leap stage**:

The transition state |Ψ'> transits (or 'jump's) between conceptual nodes in this phase. According to the Free association theory, the extent of creativity lies in the distant connections between conceptual nodes. Flatter the connection, the more divergent and novel it is. So, what makes these connections, which are spatially and/or temporally separated, happen?

To help visualize the reader, let's take the analogy of the structure of an atom [40]. Electrons move in circular or elliptical paths centering the nucleus. Each path is separated from the other via energy barriers. When the atomic structure is perturbed with an external agent, say a stream of photons (basically, energy is given to the electrons in different shells), electrons in lower energy states move to a higher energy shell, provided they receive sufficient energy to cross the barrier. Depending on the quanta of energy introduced, ground state electrons could reach any of the higher shells (sometimes even out of the atom breaking the binding energy).

Something similar, we predict, could be seen in case of divergent thinking as well. The state function |Ψ'> which constitutes the initial node, is previously acted upon by the external stimulus in form of an operator. The stimulus, we believe, is the necessary perturbation that pushes the transition state |Ψ'> to leap to the next node (the probability of leaping to the next node is equally distributed amongst all the available unless there is an introduction of contextuality by memory or interaction with environment). But what is that 'energy', similar to the atomic analogy, which is essential in this leap from primary to a secondary node?

● **Dependence on Ambiguity:**

Here, ambiguity plays a very important role. We believe that the ambiguous nature of the external stimulus provides |Ψ'> the 'energy' to leap to a secondary node. Similar phenomenon is quite common in the fields of nonlinear and quantum optics where transition or absorption rate of a particle (electron, photon) is proportional to the intensity of the perturbing light. Likewise, in this case, the nodal transition rate is dependent on the ambiguous nature of the stimulus received.

Let us use an example to clarify the point. When a person hears a word, say 'tiger', which is absolutely unambiguous in nature (i.e., the person has a well constructed knowledge or idea about the concept of 'tiger'), it is unlikely that the person would produce an unique or novel way of associating 'tiger' with, say 'suspension bridge' (provided the person has a well developed idea about it too). But, a person having an ambiguous (or vague) knowledge about both concepts has a better chance of associating them in a novel way.

The tolerance of ambiguity has been studied before as a factor of a person's creative aspect [41] [42]. According to our conjecture, it is reasonable to inspect this idea as ambiguity has an important role in the divergent production.

### 3. **Chaotic Evolution stage:**

The Third and final phase of the creative process is Chaotic evolution. Before starting this part, a brief overview of chaos theory is needed. It designates a specific class of dynamical behaviour. According to SH Kellert [43], it is "the qualitative study of unstable aperiodic behaviour in deterministic dynamical systems". 'The aperiodic' reinforces the point that the same state is never repeated twice. Chaos theory has three essential properties: firstly, they are very sensitive to initial conditions. Secondly, they can display a highly disordered behavior; and third, they are deterministic, that is they obey some laws that completely describe their motion [44]. So, where is the relevance of such a theory to our quantum leap interpretation? The phases we dealt with till now has a finite time limit set on them, namely, tens or thousands of a fraction of a second. But what happens when we try to push the limit of time gradually higher? Also, what if, instead of a single external perturbation, we had a continuous flow of stimulus to deal with? Each of these perturbations is analogous to performing a measurement on a quantum system. And (unlike classical systems) each of the interactions with environment causes some irreducible effect on the system. Quite a few number of literature in the recent past has suggested that continuous measurements on a quantum system that is evolving with time is equivalent to averaging over all the possible trajectories that the particle might have taken. Also, this kind of measurements introduces some interesting conditions on the Wigner function of the quantum system [45] [46]. The Wigner function, introduced by Wigner in 1932 [47], is a probability distribution (more technically, a quasidistribution) which helps to transform the trajectory of a quantum system or operator in phase space (in terms of its position and momentum variables, i.e., $x$ and $p$ respectively) from Hilbert space. Unlike a classical system, it is not possible to measure $x$ and $p$ of a quantum system simultaneously, thanks to Uncertainty principle. Hence, we resort to the Wigner function, described as:

$$W(x,p) = \frac{1}{h} \int e^{-\frac{ipy}{h}} \Psi\left(x + \frac{y}{2}\right) \Psi^*\left(x - \frac{y}{2}\right) dy \,,$$

Where, $\psi$ is the wavefunction, $\Psi^*$ is its complex conjugate and $x$ and $p$ are position and momentum variables and $h$ is Planck's constant. The centroid of the Wigner function is the phase space point defined by the mean values of x and p, i.e., (<x>, <p>).

For this quantum system to act as a classical one, the Wigner function needs to be 'localized', that is, its distribution needs to be sharply peaked about the phase space variables so that its evolution can be described classically in terms of these variables. The revolution of the centroid of a Wigner function over time, given in [48], follows the Ehrenfest equations:

$$<\dot{x}> = <p>/m, <\dot{p}> = <F(x, t)> = -<\frac{\partial V}{\partial x}>,$$

[where <□> is the expectation value, F(x, t) is the force and V is the potential.]

Now, for a highly localized Wigner function, the expectation value <F(x, t)> can be expanded about <x> with a Taylor expansion. Doing so, and neglecting the higher terms, readily makes the equation perfectly Newtonian:

$\langle \dot{x} \rangle$ = <p>/m, $\langle \dot{p} \rangle$ = F(<x>, t), and thereby, classical.

Since the Wigner function of an unobserved quantum system rarely remains localized, measuring the system continuously with a high rate of information extraction introduces Gaussian white noise into the evolution equation:

$\langle \dot{x} \rangle$ = <p>/m + (8k)$^{1/2} \sigma_x^2 \xi(t)$
$\langle \dot{p} \rangle$ = <F(x, t)> + (8k)$^{1/2}$ C$_{xp} \xi(t)$ ;

Where $\sigma_x^2$ = variance of x, C$_{xp}$ = co variance of x and p, and $\xi(t)$ = Gaussian white noise. This makes the localized Wigner function close to a Gaussian distribution in nature [48].

The stronger and frequent the measurement is, the more noise is introduced to the system. Careful simulation conditions reveal that the trajectory of such a noise-induced system (provided that the distribution is >> ℏ) is classical and chaotic [48] [49].

The mental state function |Ψ>, an essentially quantum system, also can be described using a Wigner distribution which undergoes similar evolutions. Continuous exposure to the external perturbations introduces the necessary noise which, in principal, gives rise to classical and chaotic behaviours (unless of course the action of the system is small compared to Planck's constant ℏ. This is unlikely, since the action of the state function is much larger as one can see from neurobiological signatures like fMRI or EEG, both in spatial and temporal domain.). We hypothesize that the above discussed chaotic trajectory is the reason why individual creativity is person specific. Evolution of the mental state, in this environment, depends heavily upon the initial conditions and exhibits disorderness. That's why even after going through the same two initial phases, creative process is found to be different for everyone. The nature of the stimulus, again, decides who will be a painter and who shall excel in music.

**OVERVIEW OF THE HYPOTHESIS**:

Here, we have proposed a hypothesis that tries to explain the creative process from its inception to its evolution. This hypothesis uses quantum physics and its unusual but highly effective approach to divide the whole process into three distinct stages. In the primary or 'Pick up' stage the mental state function (operating in a Hilbert space we call 'Mental state space') interacts with an external stimulus (denoted by an operator) and begets a new state function. This change marks the start of the creative process. In the next stage, which we denote as 'Leap' stage, this state function (primary node) leaps between the conceptual nodes further, analogous to electrons jumping from lower to higher shells absorbing external energy inside an atom. Ambiguity, similar to energy, is the active agent that makes the state function leap the conceptual nodes. During the evolution, the state function continuously interacts with external environment in the form of contexts, memories or stimulus. These interactions, keep injecting noise in the evolving Wigner distribution of the state function, ultimately localizing it and making it classical and furthermore, chaotic (subjected to strict conditions). This constitutes the third and final stage, called the 'Chaotic evolution' stage. We propose that this final stage indicates individual creativity. That is, since here the state function evolves in a disorderly chaotic manner, being highly sensitive to initial conditions and also since this stage is dominated mostly by external stimulus, hence, the nature of stimulus can affect the chaotic evolution heavily. This variation in interaction with stimulus is, we believe, the reason why every creative individual has a unique nature of creativity even after going through the first two common stages.

Here, a schematic diagram of the hypothesis is given:

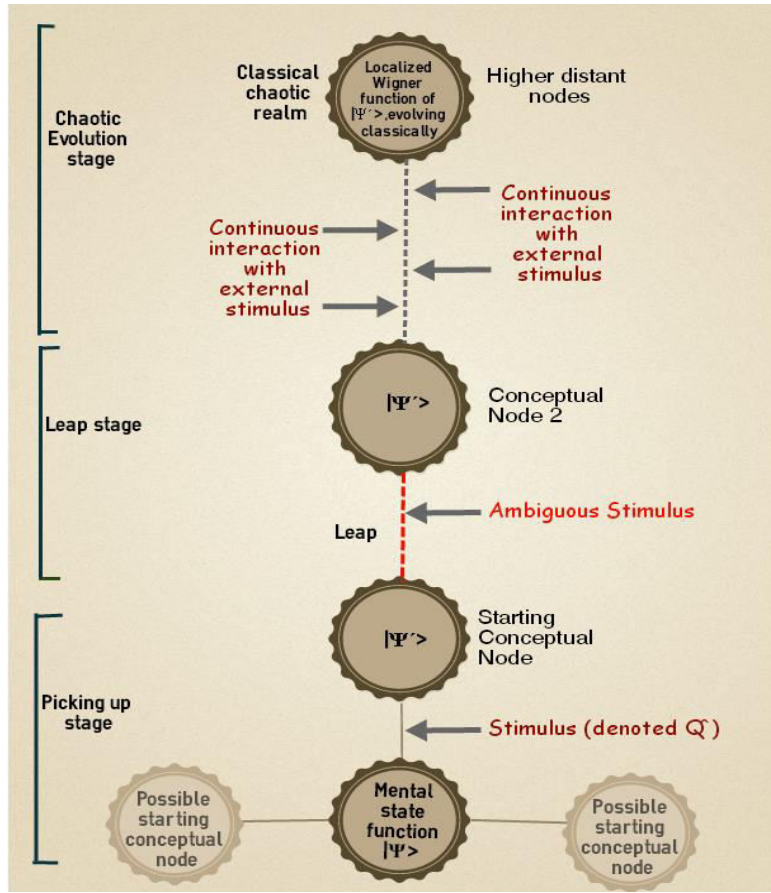

Figure 2: Schematic diagram of the Quantum Leap hypothesis

The main features and advantages of the hypothesis can be summarized as:

1. This novel hypothesis takes into account the fact that explaining creativity using classical realm can't describe the process in its entirety and hence, a quantum physical approach needs to be introduced.

2. Unlike existing ideas which are necessarily top-down (spotting the creativity, followed by the investigation of an explanation), ours is bottom-up. We try to explain the starting point of the process and discuss how it can evolve into the creative effects that one observes as a final output.

3. The effect of stimulus and its nature (degree of ambiguity, continuity) is given the utmost importance in this interpretation.

**SUGGESTED EXPERIMENTAL PROTOCOL:**

The central idea of this hypothesis lies in the fact that the nature of ambiguity of stimulus triggers the creative process. To verify the idea, the experimental protocol is to be devised such a way that the dependency of a person's creative ability on a stimulus' ambiguity is tested. The stimulus we propose to test the hypothesis here is visual. Every subject is shown a fixed number of images that are ambiguous in nature, i.e., they have more than one interpretation. They are asked to note all the interpretations of each of the figures as soon as they can see them. The transition times between the interpretations are also noted. In the next step, they are asked to sit in a software designed creative ability test which poses 40 questions to the subject containing divergent thinking, problem solving, standardized self evaluation tasks and also, ambiguous figure deciphering tasks. This software measures the creative ability of individuals using 8 different metrics. From the first part of the experiment using human response data, the necessary information can be extracted about the degree of ambiguity in stimulus, tolerance of ambiguity in individuals, and the mean transition time between different interpretations of the figures. Comparing these with the data from the creative ability test, we hope to find whether any such correlation between ambiguity and individual creative ability exists (if yes, the next step is to quantify it).

## DISCUSSION & CONCLUSION:

Further extension of this idea may be envisaged as following:
1) Development of knowledge about the creative process by detailing each of the steps.
2) Inflicting further investigation in this specific topic about the use of advanced physical theories (e.g.: quantum field theory).
3) Finding the causal relationship between Ambiguity and creativity which in turn can help promote creative aspects in young individuals.
4) In long term, encouraging 'democratic creativity': Instead of narrowing the term 'Creativity' by associating it with genius individuals, spreading it to otherwise mundane society and system will generate productivity across every social platform.
5) Improvements in the development of the skills of creativity, critical thinking and producing novel ideas are essential for developing the next generation of researchers.

Thus this paper presents a novel idea on the link between two well researched and talked about concept – Ambiguity and Creativity. The approach is based on the usage of Quantum physics, the most successful theory in the realm of the behavior of subatomic particles. The rationale behind the approach is: subatomic particles are the basic constituents of the human brain and hence their dynamics deserve detailed discussions during the investigation of any such cognitive phenomenon. The physical and mathematical ideas regarding it also approve this kind of approach- of course to be verified by different experimental protocols. The suggested ideas here can be regarded as a trigger that can enrich the knowledge of this very critical and complex domain of Creativity involving the human brain and its mechanisms.

## ACKNOWLEDGEMENT:

One of the authors, Souparno Roy, would like to acknowledge Department of Science and Technology (DST), Govt. of West Bengal for providing him Junior Research Fellowship under the project [sanction order: 860(Sanc)/ST/P/S&T/4G-3/2013] to pursue this research. Author Archi Banerjee gratefully acknowledges DST, Govt. of India for funding this work.

## REFERENCES:

[1] Sternberg RJ, Lubart TI. An investment theory of creativity and its development. Human development. 1991;34(1):1-31.
[2] Lubart TI. Product-centered self-evaluation and the creative process. Unpublished doctoral dissertation, Yale University, New Haven, CT. 1994.
[3] Sternberg RJ, Lubart TI. Defying the crowd: Cultivating creativity in a culture of conformity. Free Press; 1995.
[4] Ochse R. Before the gates of excellence: The determinants of creative genius. Cambridge University Press Archive; 1990.
[5] Ryhammar L, Brolin C. Creativity research: Historical considerations and main lines of development. Scandinavian Journal of Educational Research. 1999;43(3):259-73.
[6] Craft A. An analysis of research and literature on creativity in education. Qualifications and Curriculum Authority. 2001:1-37.
[7] Freud, S, 'Leonardo da Vinci and a memory of his childhood' in Standard Edition, Hogarth Press, London, 1957 (originally published 1910)

[8] Winnicott DW. 4. Creativity and Its Origins. Essential papers on the psychology of women. 1990 Aug 1:132.

[9] Winnicott DW. Playing and reality. Psychology Press; 1971.

[10] Galton, Francis. Hereditary genius: An inquiry into its laws and consequences. Vol. 27. Macmillan, 1869.

[11] Mednick S. The associative basis of the creative process. Psychological review. 1962;69(3):220.

[12] Guilford JP. Creativity: Yesterday, today and tomorrow. The Journal of Creative Behavior. 1967;1(1):3-14.

[13] Guilford, J.P, The nature of human intelligence, McGraw Hill, New York, NY, 1967

[14] Skinner BF. The behavior of organisms: An experimental analysis. BF Skinner Foundation; 1990.

[15] Skinner BF. Science and human behavior. Simon and Schuster; 1953.

[16] Skinner BF. Verbal behavior. BF Skinner Foundation; 2014

[17] Dietrich A, Kanso R. A review of EEG, ERP, and neuroimaging studies of creativity and insight. Psychological bulletin. 2010 ;136(5):822.

[18] Abraham A. The neuroscience of creativity: a promising or perilous enterprise?. Creativity and cognitive neuroscience. 2012:15-24.

[19] Guilford JP. Intelligence: 1965 model. American Psychologist. 1966;21(1):20.

[20] Torrance, E.P, Torrance tests of creativity, Personnel Press, Princeton, 1966

[21] Torrance EP. Predictive validity of the Torrance Tests of Creative Thinking. The Journal of Creative Behavior. 1972.

[22] Torrance EP. The nature of creativity as manifest in its testing. The nature of creativity. 1988 May 27:43-75.

[23] Collins, A.M. and Loftus, E.F., A spreading-activation theory of semantic processing. Psychological review, 1975,82(6), p.407.

[24] Freedman JL. Increasing creativity by free-association training. Journal of experimental psychology. 1965 Jan;69(1):89.

[25] Busemeyer JR, Bruza PD. Quantum models of cognition and decision. Cambridge University Press; 2012 Jul 26.

[26] Mehra, Jagdish, and Helmut Rechenberg. "The Historical Development of Quantum Mechanics, vol. 1." (1982).

[27] Resnick, R., and R. Eisberg. "Quantum physics of atoms, molecules, solids, nuclei and particles." (1985): 114-116.

[28] Bohm, D., 1990. A new theory of the relationship of mind and matter. *Philosophical psychology*, *3*(2-3), pp.271-286.

[29] Mohsen, Razavy. "Quantum Theory of tunneling." (2003).

[30] Hameroff, S. (1998). Quantum computation in brain microtubules? The Penrose-Hameroff'Orch OR'model of consciousness. *Philosophical Transactions-Royal Society of London Series A Mathematical Physical and Engineering Sciences*, 1869-1895.

[31] Monroe C, Meekhof DM, King BE, Wineland DJ. A" Schrodinger cat" superposition state of an atom. Science. 1996 May 24;272(5265):1131

[32] Braginsky, Vladimir B., Vladimir Borisovich Braginskiĭ, Farid Ya Khalili, and Kip S. Thorne. *Quantum measurement*. Cambridge University Press, 1995.

[33] Schrödinger, E. "Mathematical proceedings of the cambridge philosophical society." *Mathematical Proceedings of the Cambridge Philosophical Society*. Vol. 31. 1935.

[34] Thiagarajan, Tara C., et al. "Coherence potentials: loss-less, all-or-none network events in the cortex." *PLoS Biol* 8.1 (2010): e1000278.

[35] Fisher, Matthew PA. "Quantum cognition: the possibility of processing with nuclear spins in the brain." *Annals of Physics* 362 (2015): 593-602.

[36] Arndt M, Juffmann T, Vedral V. Quantum physics meets biology. HFSP journal. 2009;3(6):386-400.

[37] Tarlaci S. Why we need quantum physics for cognitive neuroscience. NeuroQuantology. 2010;8(1).

[38] Robertson HP. The uncertainty principle. Physical Review. 1929 Jul 1;34(1):163.

[39] Manousakis E. Quantum formalism to describe binocular rivalry. Biosystems. 2009 Nov 30;98(2):57-66.

[40] Martin RM. Electronic structure: basic theory and practical methods. Cambridge university press; 2004 Apr 8.

[41] Zenasni F, Besançon M, Lubart T. Creativity and tolerance of ambiguity: An empirical study. The Journal of Creative Behavior. 2008;42(1):61-73.

[42] Furnham A, Ribchester T. Tolerance of ambiguity: A review of the concept, its measurement and applications. Current psychology. 1995;14(3):179-99.

[43] Kellert SH. In the wake of chaos: Unpredictable order in dynamical systems. University of Chicago press; 1994

[44] Faure P, Korn H. Is there chaos in the brain? I. Concepts of nonlinear dynamics and methods of investigation. Comptes Rendus de l'Académie des Sciences-Series III-Sciences de la Vie. 2001;324(9):773-93.

[45] Bhattacharya T, Habib S, Jacobs K. Continuous quantum measurement and the quantum to classical transition. Physical Review A. 2003;67(4):042103.

[46] Habib S, Bhattacharya T, Greenbaum B, Jacobs K, Shizume K, Sundaram B. Chaos and quantum mechanics. Annals of the New York Academy of Sciences. 2005;1045(1):308-32..

[47] Wigner E. On the quantum correction for thermodynamic equilibrium. Physical review. 1932;40(5):749.

[48] Bhattacharya T, Habib S, Jacobs K. The emergence of classical dynamics in a quantum world. arXiv preprint quant-ph/0407096. 2004 Jul 14.

[49] Bhattacharya T, Habib S, Jacobs K. Continuous quantum measurement and the emergence of classical chaos. Physical review letters. 2000;85(23):4852.